\documentclass[pra,letterpaper,10pt,twocolumn]{revtex4}
\usepackage{graphicx}
\usepackage[final,dvips]{epsfig}
\usepackage{color}
\usepackage{dcolumn}  % Align table columns on decimal point
\usepackage{bm}       % bold math
\usepackage{array}    % table making
\usepackage{amssymb,amsmath}
\usepackage[colorlinks,citecolor=red]{hyperref}
\emergencystretch=\maxdimen
\include{MyCommand}

\begin{document}

\title{Interplay of Magnetic Order, Pairing and Phase Separation in a One Dimensional Spin Fermion Model}

\author{Wenjian Hu}
\author{Richard T. Scalettar}
\author{Rajiv R.P. Singh}
\affiliation{
Department of Physics, University of California Davis, 95616 CA  USA
}

\begin{abstract}
We consider a lattice model of itinerant electrons coupled to an array
of localized classical Heisenberg spins. The nature of the ground state
ordered magnetic phases that result from the indirect spin-spin coupling
mediated by the electrons is determined as a function of density and the
spin-fermion coupling $J$.  At a fixed chemical potential, spiral phases
exist only up to values of $J$ which are less than roughly half the
electronic bandwidth. At a fixed electron density and near half filling,
the system phase separates into a half-filled antiferromagnetic phase
and a spiral phase. The ferromagnetic phases are shown to be fully
polarized, while the spiral phases have equal admixture of up and down
spins.  Phase separation survives in the presence of weak pairing field
$\Delta$ but disappears when $\Delta$ exceeds a critical value
$\Delta_c$. If pairing fields are large enough, an additional spiral
state arises at strong coupling $J$.  The relevance of this study,
especially the phase separation, to artificially engineered systems of
adjacent itinerant electrons and localized spins is discussed. In
particular, we propose a method which might allow for the braiding of
Majorana fermions by changing the density and moving their location as
they are pulled along by a phase separation boundary.
\end{abstract}

\maketitle

%%%%%%%%%%%%%%%%%%%%%%%%%%%%%%%%%%%%%%%%%%%%%%%%%%%%%%%%%%%%%%%%%%%%%%%%%%
\section{Introduction}
%%%%%%%%%%%%%%%%%%%%%%%%%%%%%%%%%%%%%%%%%%%%%%%%%%%%%%%%%%%%%%%%%%%%%%%%%%

Many-body effects in solids are commonly explored either by Hamiltonians
which include explicit electron-electron interactions, {\it e.g.~}the
Hubbard or Periodic Anderson models, or else by Hamiltonians containing
interactions of free electrons with other (quantum spin or phonon)
degrees of freedom, {\it e.g.~}the Kondo or Holstein models,
respectively.  The distinction between these two descriptions is not
completely sharp-- the Kondo model is a large $U$ limit of the Periodic
Anderson Model, for example.  Even more generally, the introduction of
auxiliary fields allows the mapping of Hamiltonians with direct
electron-electron interactions to free fermions coupled to a
`Hubbard-Stratonovich' field.  In all these cases, the quantum spin,
phonon, or Hubbard-Stratonovich variables depend on both space and
imaginary time.

Suppressing the imaginary time dependence represents an approximation to
including the full effect of quantum fluctuations.  Nevertheless, many
of the interesting effects of electronic interactions can still be
studied in this regime.  This approximation has been motivated, for
example, in cases where large (``classical") core spins are coupled to
an itinerant electron band, a situation which occurs in the double
perovskite Sr$_2$FeMoO$_6$ where the 3d$^5$ configuration of Fe$^{3+}$
forms a spin-5/2 with which the Mo electron interacts
\cite{Erten11,Sanyal09,Meetei13}.  Similarly, applications of classical
spin-fermion models to a broad class of materials including
nickelates\cite{Johnston14,Park12}, cuprate
superconductors\cite{Buhler00}, and iron
superconductors\cite{Yin10,Lv10,Liang13}, have been reported, with the
elucidation of subtle many body effects including magnetic and charge
domain formation, and site-selective Mott transitions. In these
materials, the coupling of an itinerant electron band to classical
degrees of freedom is argued to capture the quantum mechanics of fast
electronic motion in contact with the slower degrees of freedom, much in
the spirit of the Born-Oppenheimer approximation or the Car-Parrinello
method\cite{Car85}.

In a technically similar spirit of treating interactions by a simplified
time independent variable that leaves the electronic problem quadratic,
proximity induced superconductivity can be described by a pairing field
bilinear in fermion operators\cite{Kitaev01}. The combined effects of
frozen spin and pairing configurations lead to many interesting features
including topological superconductivity and Majorana
states\cite{Sau12,Alicea12}. 

The realizations of spin-fermion models described above have largely
concerned tight binding Hamiltonians where the electrons move on a
lattice. Very recently, the approach was also applied to a gas of
fermions moving in the continuum in one dimension, but coupled to a
regular array of classical spins\cite{Schecter15}. 
The itinerant electrons mediate a Ruderman-Kittel-Kasuya-Yoshida
(RKKY)\cite{Ruderman54,Kasuya56,Yosida57} interaction. The goal of the
work was to determine the ground state phase diagram in the plane of the
exchange coupling between the classical and electron spins and the
electron density, and explore the possibility of spiral spin states
driving one dimensional superconductors into topological
phases\cite{Pientka13,Pientka14,Braunecker13,Klinovaja13}.

In this paper we study the lattice version of this problem (See also
\cite{Vazifeh13}).  We show that, similar to the continuum case,
ferromagnetism (F) occurs at low filling and is replaced by  a spiral
configuration of classical spins, which minimizes the energy only when
the exchange constant $J$ is smaller than roughly half the bandwidth W.
The spiral phase gives way to commensurate antiferromagnetic (AF) order
as $J$ increases.  One central observation of our work is that near
half-filling and at a fixed electron density, the system phase separates
into a half-filled AF phase and a spiral phase of reduced (less than
half-filling) or enhanced (greater than half-filling) density at weak
coupling $J\lesssim W/2 = 2t$.  We show that this spiral-AF phase
separation initially survives in the presence of a weak pairing field
$\Delta$, but for larger $\Delta$ the transition becomes continuous.  We
observe that the effect of $\Delta$ is richer at strong coupling $J$.
For $\Delta$ small the transition from AF to F is direct, without an
intervening spiral phase, and exhibits phase separation.  With
increasing $\Delta$, a spiral phase emerges.   The spiral to AF
transition is initially discontinuous, but becomes continuous for larger
values of $\Delta$.  We discuss the implication of these results for an
artificially engineered system of adjacent itinerant electrons and
localized spins.

%%%%%%%%%%%%%%%%%%%%%%%%%%%%%%%%%%%%%%%%%%%%%%%%%%%%%%%%%%%%%%%%%%%%%%%%%%
\section{Model and Methods}
%%%%%%%%%%%%%%%%%%%%%%%%%%%%%%%%%%%%%%%%%%%%%%%%%%%%%%%%%%%%%%%%%%%%%%%%%%

We first consider a lattice of one dimensional itinerant electrons coupled to
classical Heisenberg spins $\vec S_l$ in the Grand Canonical
Ensemble (GCE),
\begin{align}
H&=-t \sum_{l\sigma} ( c_{l+1\,\sigma}^{\dagger}
c_{l\sigma}^{\phantom{\dagger}}
 + c_{l\sigma}^{\dagger}c_{l+1\,\sigma}^{\phantom{\dagger}})
\nonumber \\
& -\mu\sum_{l\sigma} n_{l\sigma}
+J \sum_{l}  \vec s_l \cdot \vec S_l
\label{eq:ham}
\end{align}
Here the fermionic spin components are
$s_{lx}^{\phantom{\dagger}}
=  c_{l\uparrow}^{\dagger}c_{l\downarrow}^{\phantom{\dagger}}
+  c_{l\downarrow}^{\dagger}c_{l\uparrow}^{\phantom{\dagger}};
\hskip0.04in
s_{ly}^{\phantom{\dagger}}
=  -i \, c_{l\uparrow}^{\dagger}c_{l\downarrow}^{\phantom{\dagger}}
   +i \,  c_{l\downarrow}^{\dagger}c_{l\uparrow}^{\phantom{\dagger}};$
and
$s_{lz}^{\phantom{\dagger}}
=  c_{l\uparrow}^{\dagger}c_{l\uparrow}^{\phantom{\dagger}}
  - c_{l\downarrow}^{\dagger}c_{l\downarrow}^{\phantom{\dagger}}
           $.

A single spin in a conduction electron sea induces a magnetic
polarization which can then couple to other spins, so that the itinerant
electrons mediate an RKKY interaction.  At $T=0$ this interaction falls
off as $1/r^d$ in $d$ dimensions, and oscillates in phase with the Fermi
wave vector $k_F$.  A classical Ising model in $d=1$ with (unfrustrated)
power law interactions $J(r) \sim 1/r^{p}$ has no long range order at
finite $T$ if $p>2$, but orders, with mean field exponents, for $1 < p <
3/2$. Order at finite $T$ with continuously varying exponent occurs in
between these cases $3/2 < p < 2$.  For $p<1$ ordering occurs at all
$T$.  1D classical Heisenberg spins with power law interactions also
have long range order at finite temperature for $1<p<2$ but no
transition for $p \geqslant 2$\cite{Cavallo02}, which suggests that, in
principle, the $1/r$ RKKY long range interaction could result in spin
ordering at $T>0$ in one dimension.  However, for finite temperatures
the power law is multiplied by a decaying exponential $e^{-r/\xi(T)}$,
with correlation length $\xi(T)$ diverging at $T=0$.  Thus one expects
only (possibly rapid) cross-overs to quasi-ordered phases at $T \neq 0$.

Since $H$ is quadratic in the fermion creation and destruction
operators, it may be solved for an arbitrary Heisenberg spin
configuration $\{ \vec S_l \}$ by diagonalizing a matrix of dimension
$2L$, where $L$ is the number of sites in the chain. The factor of two
arises from the mixing of the fermionic spin components through their
coupling to $\{ \vec S_l\}$. The competition between the
Fermi vector $k_F$, which depends on doping, and the AF exchange, leads
to a spiral phase\cite{Braunecker09,Braunecker0902}, where the classical
Heisenberg spins form a spiral magnetic structure. We do not consider
arbitrary Heisenberg spin configurations $\{ \vec S_l \}$ and instead
make an {\it ansatz} of a spiral configuration\cite{Vazifeh13}.  The
most general such form is $\vec{S_l}=\hat{x} \cos(q_1 l) \cos(q_2 l +
\phi) +\hat{y} \cos(q_1 l) \sin(q_2 l +\phi)+\hat{z}\sin(q_1 l)$, which,
in zero external field, should further  simplify to a planar form $ \vec
S_l  = \hat x \, {\rm cos} (ql) + \hat y \, {\rm sin}(ql)$. Here allowed
$q$ values with periodic boundary conditions are $q=n*\frac{2\pi}{L},
n=0,1,2,3......,(L-1)$.  Because of the rotational symmetry of the
fermionic part of the Hamiltonian, the properties are equivalent with
other choices for the spin plane.  Likewise, symmetry dictates that $q$
and $(2\pi-q)$ yield the same physics. 

While the full $2L$ dimensional matrix must be considered to solve
Eq.~\ref{eq:ham} in the case of arbitrary $\{S_l\}$ through a full
matrix diagonalization, with this planar, spiral {\it ansatz}, the
Hamiltonian can be solved in momentum space:
$c^{\dagger}_{l\sigma}=(1/\sqrt{N}) \sum_{k} e^{-i
kl}c^{\dagger}_{k\sigma}$.  For a specific $q$, spin $\uparrow$ fermions
of momentum $k-q$ are mixed only with spin $\downarrow$ fermions of
momentum $k$ and the Hamiltonian $H$ becomes a sum of independent 2x2
blocks:
\begin{align}
H_q &=  \sum_{k}[
-2t \,(\, {\rm cos}(k-q)  \,
 c_{k-q \, \uparrow}^{\dagger} c_{k-q\uparrow}^{\phantom{\dagger}}
+ {\rm cos}(k)  \,
 c_{k \, \downarrow}^{\dagger} c_{k\downarrow}^{\phantom{\dagger}} \,)
\nonumber \\
+&J \, (\, 
c_{k-q \, \uparrow}^{\dagger} c_{k\downarrow}^{\phantom{\dagger}}
 + c_{k\downarrow}^{\dagger} c_{k-q \, \uparrow}^{\phantom{\dagger}}
\, ) 
%% \nonumber \\
- \mu
 ( \, c_{k-q \, \uparrow}^{\dagger} c_{k-q\uparrow}^{\phantom{\dagger}}
 + c_{k \, \downarrow}^{\dagger} c_{k\downarrow}^{\phantom{\dagger}} \,)
\nonumber] \\
\label{eq:hamq}
\end{align}
This Hamiltonian produces two energy bands,
\begin{align}
E_\pm=-t(\cos k+ \cos (k-q)) -\mu
\nonumber \\
\pm \sqrt{{[t(\cos k - \cos (k-q))]}^2+{J}^2},
\end{align}
We select $t=1$ to set our energy scale.

As reviewed in the introduction, it is of interest to extend the model in Eq.~\ref{eq:ham} to include $\Delta$ in the Hamiltonian:
\begin{align}
H_{\Delta}=-t \sum_{l \sigma}(c^{\dagger}_{l+1 \sigma}c^{\phantom{\dagger}}_{l \sigma}+c^{\dagger}_{l \sigma}c^{\phantom{\dagger}}_{l+1 \sigma})-\mu \sum_{l \sigma}n_{l \sigma} \nonumber \\
 +J \sum_l \vec{s_l} \cdot \vec{S_l}+\sum_l(\Delta c^{\dagger}_{l \uparrow}c^{\dagger}_{l \downarrow} + h.c.)
 \label{eq:ham_Delta}
\end{align}
Part of the results in this paper will be to determine the effect of
$\Delta$ on the phase diagram, specifically, phase separation.  To find
the ground state energy, the Hamiltonian can again be transformed to
momentum space.  The combination of the spin up/down mixing due to J and
the pairing yields a 4x4 structure,
\begin{align}
H_{\Delta}=\frac{1}{2}\sum_k
  \vec{v_k}^{\dagger}
  M_k \vec{v_k}
  +\frac{1}{2}\sum_k[2J+\epsilon_{-(k-q)}+\epsilon_{-k}]
\end{align}
where $\epsilon_{k-q}=\epsilon_{-(k-q)}=-2t\cos(k-q)-\mu$, 
$\epsilon_k=\epsilon_{-k}=-2t\cos(k)-\mu$, $\vec{v_k}^{\dagger}=
\left[ {\begin{array}{cccc}
   c^{\dagger}_{k-q,\uparrow} & c^{\dagger}_{k,\downarrow} & c^{\phantom{\dagger}}_{-(k-q),\downarrow} & c^{\phantom{\dagger}}_{-k,\uparrow}\\
  \end{array} } \right]$, and 
\begin{align}
M_k=\left[ {\begin{array}{cccc}
   \epsilon_{k-q} & J & \Delta & 0\\
   J & \epsilon_k & 0 & \Delta\\
   \Delta & 0 & -\epsilon_{-(k-q)} & -J\\
   0 & \Delta & -J & -\epsilon_{-k}\\
  \end{array} } \right].
  \label{eq:energy_matrix}
\end{align}
$M_k$ has four eigenvalues, $\lambda_1$, $\lambda_2$, $\lambda_3$, $\lambda_4$ ($\lambda_1>\lambda_2>0>\lambda_3>\lambda_4$) and the corresponding normalized eigenvectors $\vec{v_1}$,$\vec{v_2}$,$\vec{v_3}$,$\vec{v_4}$. We define a new set of operators:
\begin{align}
\left[ {\begin{array}{c}
   \gamma^{\phantom{\dagger}}_{k,1} \\ \gamma^{\phantom{\dagger}}_{k,2} \\ \gamma^{\dagger}_{k,3} \\ \gamma^{\dagger}_{k,4}\\
  \end{array} } \right]=
   S_k^T
  \left[ {\begin{array}{c}
   c^{\phantom{\dagger}}_{k-q,\uparrow} \\ c^{\phantom{\dagger}}_{k,\downarrow} \\ c^{\dagger}_{-(k-q),\downarrow} \\ c^{\dagger}_{-k,\uparrow}\\
  \end{array} } \right]
  \label{eq:transformation}
\end{align}
where the matrix $S_k=[\vec{v_1},\vec{v_2},\vec{v_3},\vec{v_4}]$. 
In terms of these new canonical operators,
\begin{align}
H_{\Delta}=\frac{1}{2} \sum_k[\lambda_1 \gamma^{\dagger}_{k,1} \gamma^{\phantom{\dagger}}_{k,1}+\lambda_2 \gamma^{\dagger}_{k,2} \gamma^{\phantom{\dagger}}_{k,2}-\lambda_3 \gamma^{\dagger}_{k,3} \gamma^{\phantom{\dagger}}_{k,3} \nonumber \\
-\lambda_4 \gamma^{\dagger}_{k,4} \gamma^{\phantom{\dagger}}_{k,4}
+\lambda_3+\lambda_4+\epsilon_{k}+\epsilon_{k-q}]
\end{align}
Since the ground state has no excited quasi-particles, the ground state energy is:
\begin{align}
E_{0}=\frac{1}{2} \sum_k[\lambda_3(k)+\lambda_4(k)+\epsilon_{k}+\epsilon_{k-q}] \nonumber \\
=\frac{1}{2} \sum_k[\lambda_3(k)+\lambda_4(k)]-L\mu
\label{eq:energy_ground}
\end{align}
%%%%%%%%%%%%%%%%%%%%%%%%%%%%%%%%%%%%%%%%%%%%%%%%%%%%%%%%%%%%%%%%%%%%%%%%%%
\section{Magnetic Phase Diagram}
%%%%%%%%%%%%%%%%%%%%%%%%%%%%%%%%%%%%%%%%%%%%%%%%%%%%%%%%%%%%%%%%%%%%%%%%%%

\begin{figure}[!h]
      \includegraphics[width=0.98\columnwidth]{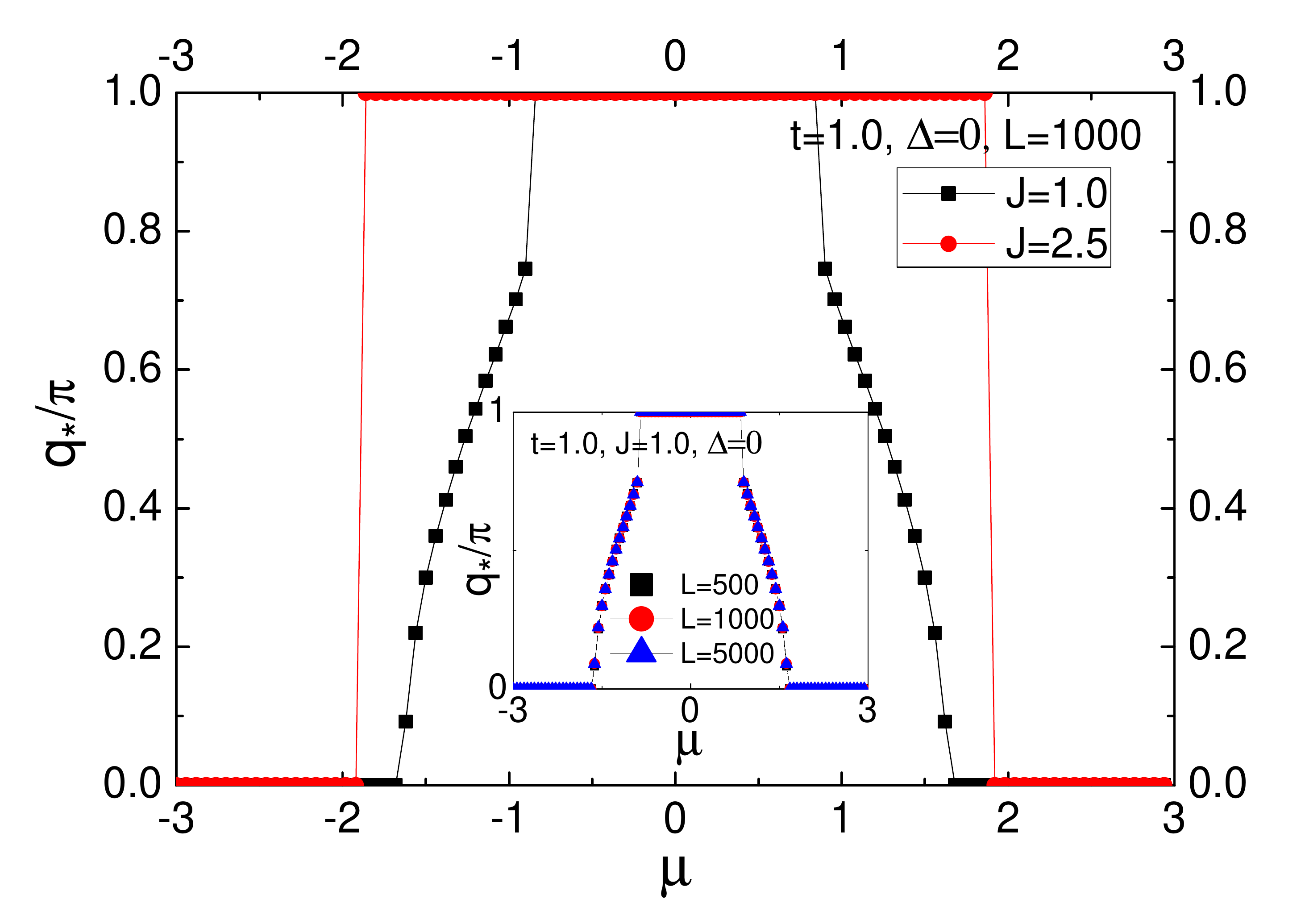} 
   \caption{
Classical spin wave vector $q_*$ which minimizes the ground state energy, as a function of chemical potential $\mu$ for fixed $J=1.0$ and $J=2.5$. \underline{Inset:}
Red circles denote results with lattice size $L=1000$ sites. Data for $L=500$
(black squares) and $L=5000$ (blue triangles) indicate finite size effects are small.
      \label{fig:q_vs_mu}
  }
   \end{figure}

We first consider $\Delta=0$. For each $J$ we determine the optimal
ordering wave vector $q_*$ by minimizing the ground state energy,
obtained by summing all eigenenergies of Eq.~\ref{eq:hamq} up to the
desired chemical potential $\mu$.  Typical results are shown in
Fig.~\ref{fig:q_vs_mu}.  We see that F order is preferred at low
density, and gives way first to spiral and then AF order as $\mu$
increases at fixed $J=1$. At $J=2.5$, there is a direct phase transition
from F to AF.  The Hamiltonian Eq.~\ref{eq:ham} is particle-hole
symmetric, so the optimal $q$ is the same for $-\mu$ and $+\mu$.

\begin{figure}[!h]
      \includegraphics[width=0.98\columnwidth]{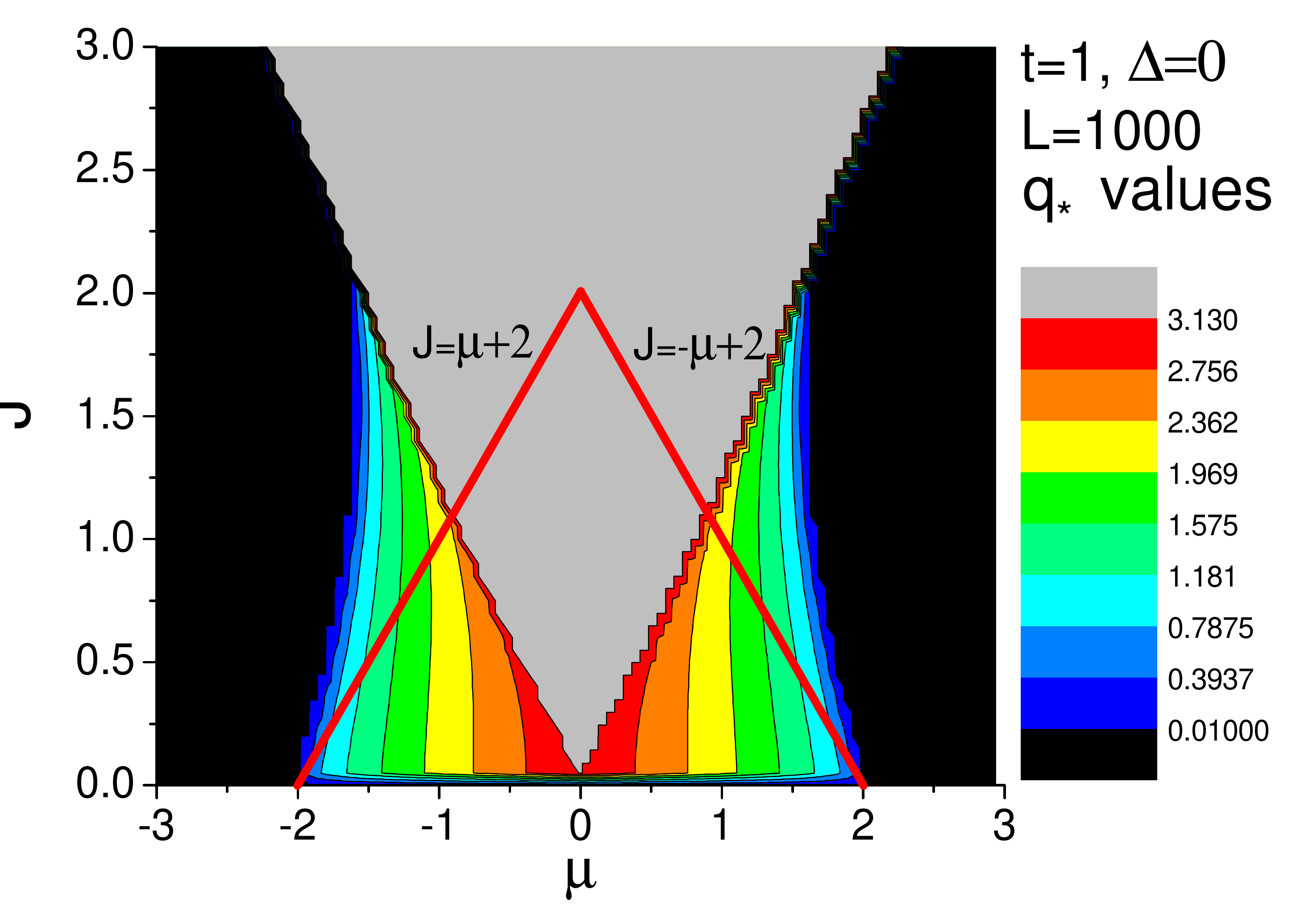} 
   \caption{
      \label{fig:phase_diagram}
Phase diagram in the chemical potential ($\mu$), exchange constant ($J$)
plane.  At low and high densities (the Hamiltonian
is particle-hole symmetric) ferromagnetic order
$q_*=0$ minimizes the energy.  For $J \lesssim W/2 = 2t$, F order
gives way first to spiral (incommensurate
$q_*$) and then AF ($q_*=\pi$) order.
The red lines mark the boundary between full and partial
polarization in a F state.  See text.
  }
   \end{figure}

Performing the calculation in Fig.~\ref{fig:q_vs_mu} for different $J$
generates the ground state phase diagram of
Fig.~\ref{fig:phase_diagram}.  The topology is qualitatively similar to
the continuum reported in Ref.~\cite{Schecter15}.  The $J=0$ corners of
the spiral phase are at $\mu= \pm 2t$ where one enters/leaves the band. The precise value $J_c$ above which the spiral phase
vanishes must be determined numerically.  However, we expect $J_c$ to
the same order as $t$, as is the case in the continuum
model\cite{Schecter15} where $J_c \sim \sqrt{\mu/m}$ where $\mu$ is the
chemical potential and $m$ is the effective mass, since in the lattice
model $\mu \sim t$ and $m \sim 1/t$. In particular, for all chemical
potentials, spiral order gives way to AF order as $J$ increases. A
difference is the almost linear AF phase boundaries we find here.  This
near linearity is a consequence of the dispersion relation in the AF
phase on a lattice, $E_{{\rm AF} \pm} = \pm \sqrt{J^2 + 4 t^2 {\rm
cos}^2 k}$, which gives a gap $\Delta=2J$. When the
chemical potential $\mu$ is in the gap $-J<\mu<J$, the density is
pinned at
half-filling, where the AF phase dominates. 
Thus, in the absence of phase separation, we
expect linear AF phase boundaries. As described further below,
however, this argument must be refined because of the occurrence of
phase separation.  Another distinction from the continuum model is the
particle-hole symmetry of Eq.~\ref{eq:ham}.  In the continuum system,
$\rho$ can extend to arbitrarily large values, as opposed to the maximal
density $\rho=2$ fermions per site in the lattice case. 

One can ask whether the F order is fully or partially polarized, By "fully polarized" we mean that either the majority spin density
equals the total density $\rho$ and the minority spin density vanishes
(below half filling), or the majority spin density is saturated at unity
and the minority spin density is $\rho-1$ (above half filling). That is, 
the system is as polarized as possible, consistent with its overall
particle density. Because our spiral state
is in the xy plane, we consider $N_{x\sigma}=\sum_l \langle
c^{\dagger}_{l,x\sigma}c^{\phantom{\dagger}}_{l,x\sigma}\rangle$, where
$c_{l,x\sigma}=(c_{l\uparrow}+\sigma c_{l\downarrow})/\sqrt{2}$ in terms
of the operators $c_{l\uparrow}$ and $c_{l\downarrow}$ which create
fermions with spin up/down in the $z$ direction at site $l$. At $q_*=0$
the eigenspectrum is $E_{{\rm F}\pm}=-2t \, {\rm cos} k  - \mu \pm J$.
For small $J$, the upper band $E_+$ will begin to be occupied for
relatively low chemical potential (density).  As $J$ increases, full
polarization persists to larger chemical potential (density).  The red
lines in Fig.~\ref{fig:phase_diagram} give the values of $J$ above which
partial polarization occurs.  These lie entirely in the spiral portion
of the phase diagram, so that we conclude only fully polarized F phases
occur.

 \begin{figure}[!h]
      \includegraphics[width=0.98\columnwidth]{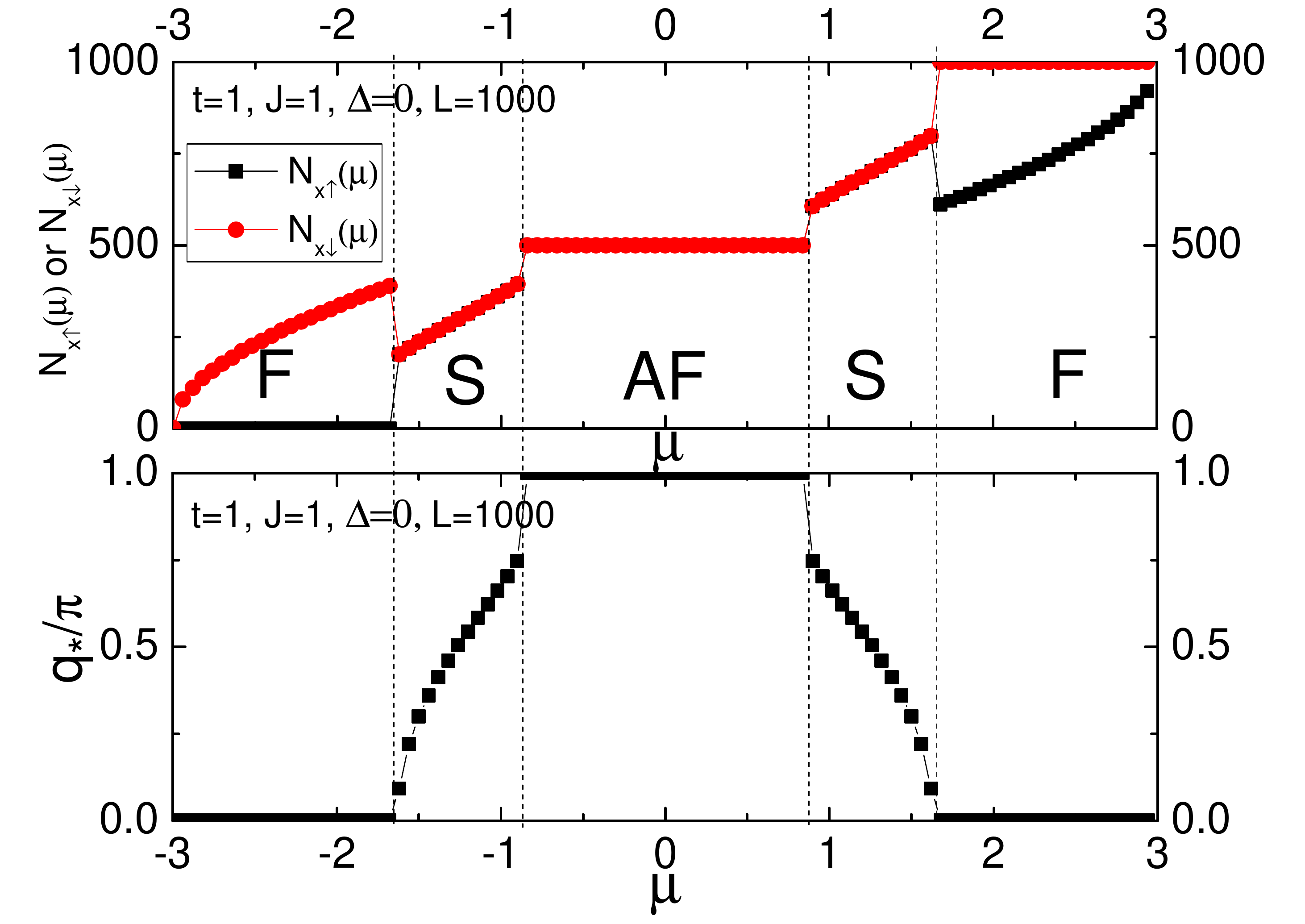} 
   \caption{
      \label{fig:Nx_up_dn}
\underbar{Top:}
Populations of the individual species with spin in the $\pm \hat x$
directions, as functions of chemical potential. The F
phase at low (and high) density is fully polarized, while the spiral and
AF phases have balanced populations.
\underbar{Bottom:} 
The corresponding ground state(F, S or AF). Discontinuities in $N_{x\sigma}$ occur precisely at the magnetic phase boundaries.}
\end{figure}

Further details into the magnetic phases can be obtained by separately
computing the densities of the two spin species as functions of chemical
potential.  This is shown in Fig.~\ref{fig:Nx_up_dn}. The vertical lines
are the magnetic boundaries, which are perfect aligned with abrupt
changes in $N_{x\sigma}$.  The fully polarized nature of the F phase is
emphasized by the fact that the density of one species is zero below
entrance into the spiral phase, at which point the two species become
equally populated.

   \begin{figure}[!h]
      \includegraphics[width=0.98\columnwidth]{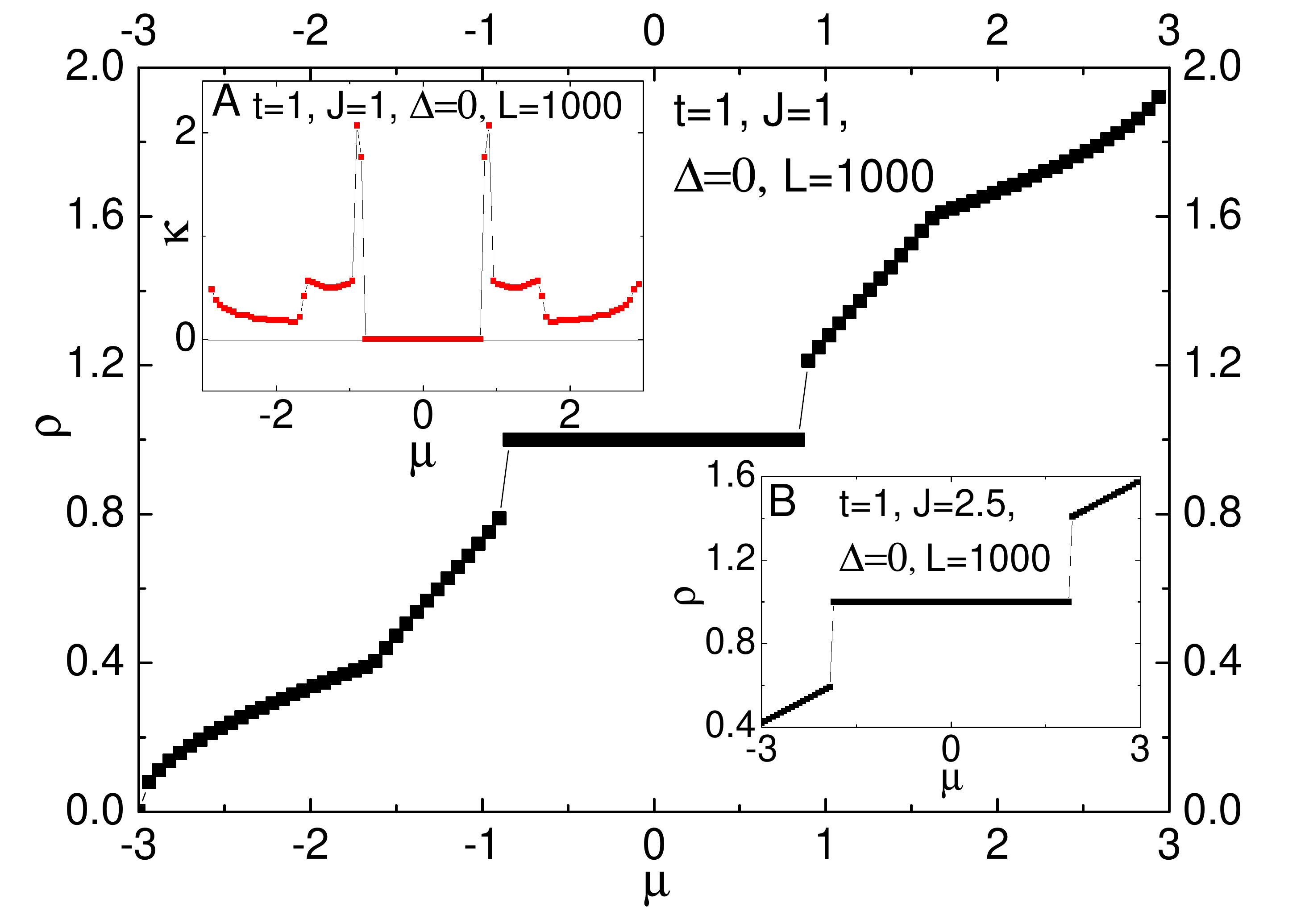}
   \caption{
      \label{fig:rho_vs_mu}
Density as a function of chemical potential at $J=1$.   The F-spiral phase boundary is signalled by a jump in the compressibility $\kappa = d\rho/d\mu$. The density jumps abruptly at the spiral-AF boundary. \underline{Inset A:} Compressibility $\kappa$ vs $\mu$. \underline{Inset B:} Density as a function of $\mu$ at $J=2.5$.
  }
   \end{figure}

Fig.~\ref{fig:rho_vs_mu} shows the density as a function of chemical
potential for a cut across the phase boundary at $J=1$.  There is a kink
in $\rho(\mu)$ at $\mu \approx -1.63$ where the compressibility $\kappa
= d\rho/d\mu$ jumps upon entering the spiral phase from the ferromagnet.
$\rho$ is discontinuous at the entry to the AF phase from spiral order.
The plateau at $\rho=1$ extends over a range of $\mu$ roughly given by
the AF gap $\Delta=2J$ at $q_*=\pi$.  (See above.) For the case $J=2.5$,
density $\rho$ is also discontinuous. The plateau range at $\rho=1$ is
somewhat less than $2J$, which is shown in the inset B of
Fig.~\ref{fig:rho_vs_mu}. Entrance into the AF phase is from the F phase
rather than the spiral phase.

   \begin{figure}[!h]
      \includegraphics[width=0.98\columnwidth]{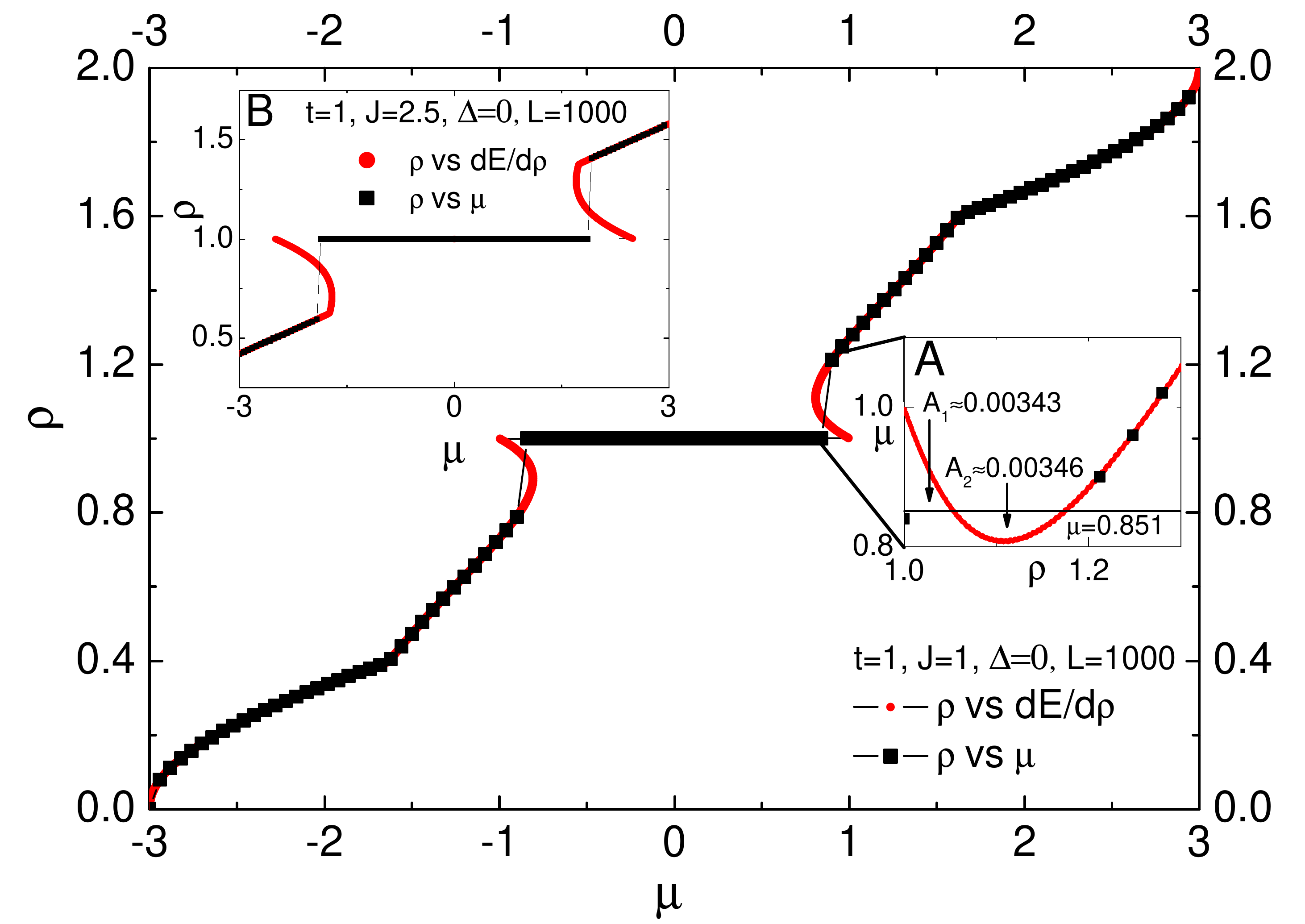}
   \caption{
Density is shown as a function of chemical potential in the graph.
Black squares denote $\rho$ vs $\mu$ in the GCE.
Red circles denote $\rho$ vs $\mu=dE_0/d\rho$ in
the CE. 
\underbar{Inset A:} Enlarged $\mu$ vs $\rho$ graph. 
The two areas $A_1\approx A_2$
satisfy the Maxwell construction.  Here the
critical $\mu$ is 0.851. 
\underbar{Inset B}: $\rho$ vs $\mu$ in GCE and $\rho$ vs $dE/d\rho$ in CE at $J=2.5$, which shows a direct phase transition from F to AF.
      \label{fig:rho_vs_dE_drho}
  }
   \end{figure}

The discontinuity in $q_*$ at the spiral-AF boundary and the F-AF
boundary seen in Fig.~\ref{fig:q_vs_mu} indicates the presence of
spiral-AF phase separation and F-AF phase separation, which is
consistent with the discontinuity in $\rho$ in Fig.~\ref{fig:rho_vs_mu}.
This presence of this first order phase transition agrees with the
continuum case\cite{Schecter15}. The case of spiral-AF phase
separation is further discussed below. At $J=1$ there is a
thermodynamically unstable range of densities $0.823 \lesssim \rho < 1$
where separate AF and spiral domains coexist.  This can be further
probed by working in the Canonical Ensemble (CE) and computing the
chemical potential via a finite difference $\mu = E_0(N+1) - E_0(N)$.
Here $E_0$ denotes the ground state energy per site.  As $N$ increases
past a critical value, $\mu$ begins to decrease so that $\kappa = d
\rho/d\mu < 0$.  Phrased alternatively, the ground state energy is
concave down, $d^2 E_0/d\rho^2 < 0$.  This indicates the boundary of the
region of phase separation around half-filling.  For a lattice with
those values of $\rho$, the energy can be lowered by phase separating
into distinct spiral and AF regions.  We have verified that the Maxwell
equal-area construction is satisfied, which as is expected at the
spiral-AF first order phase transition.

The main panel of Fig.~\ref{fig:rho_vs_dE_drho} gives $\rho$ as a
function of $\mu$ for both the CE and GCE at $J=1$.  The negative
curvature of $E_0(\rho)$  in the CE is reflected in the bending back of
the CE curve for $\rho(\mu)$.  This signature of phase separation at the
spiral to AF boundary also occurs at the F to AF boundary, as shown for
$J=2.5$ in inset B of Fig.~\ref{fig:rho_vs_dE_drho}.

To probe the details of coexistence further, we verified that
the fractions $f$ $(1-f)$ of the chain in the spiral (AF) phases obey
\begin{align}
\rho=\rho_S*f+\rho_{AF}*(1-f)
\end{align}
Where $\rho$ is the overall mixture density, $\rho_{AF}$ equals $1.0$
and $\rho_S(J)$ is the spiral state density in the mixture. For example,
at $J=1$, $\rho_S=0.823$.  This relation is also obeyed for phase
separation at the F-AF boundary, with $\rho_S$ replaced by $\rho_F$.
For $J=2.5$ we find $\rho_F=0.598$.

\begin{figure}[!h]
      \includegraphics[width=0.98\columnwidth]{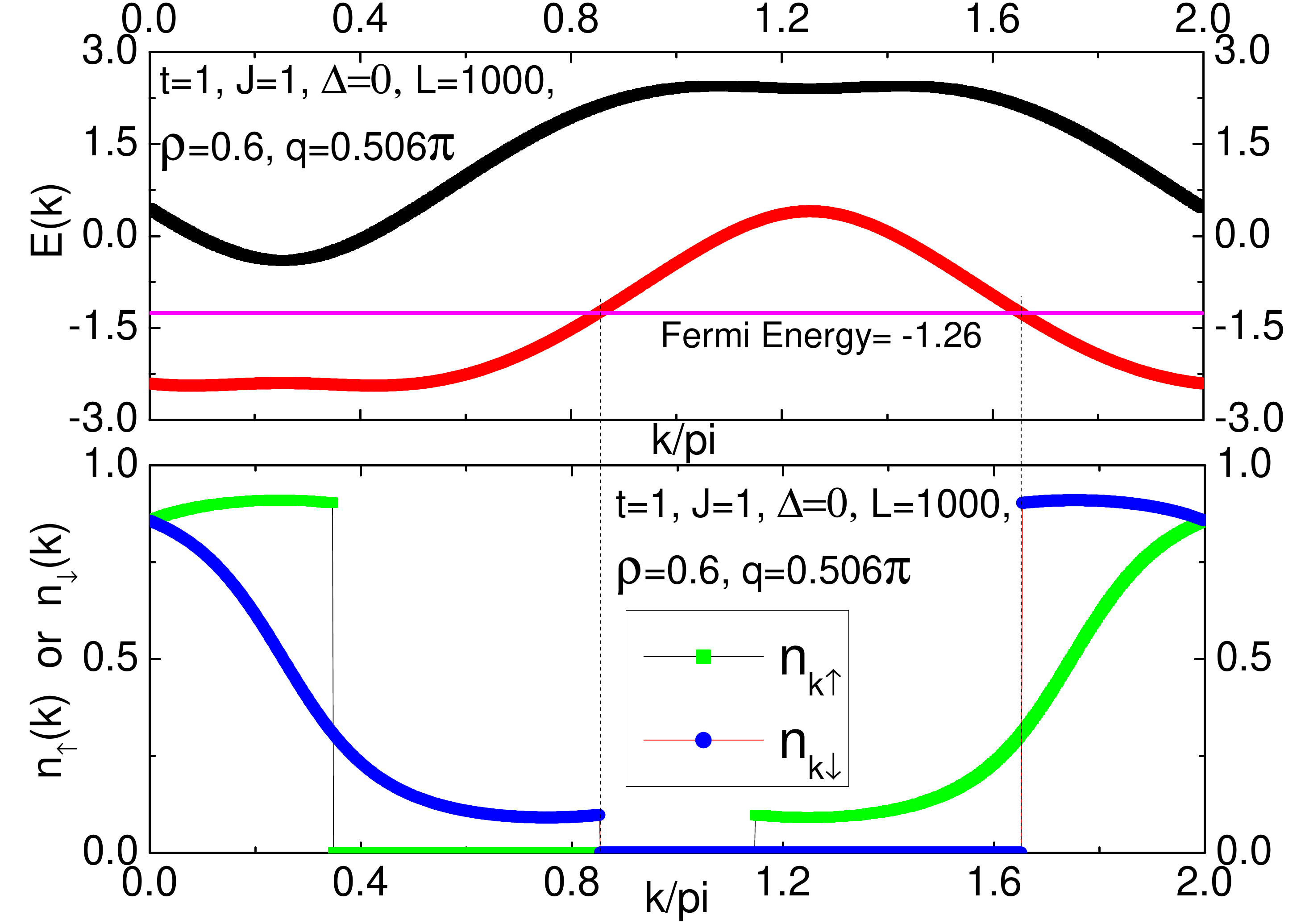} 
   \caption{
\underbar{Top:}
The mixing of $k$ with $k-q$, for $q=0.506\pi$,
gives rise to a pair of overlapping
energy bands.
\underbar{Bottom:}
Occupation numbers in the k-space. The jumps of $n_{\uparrow}(k)$ or
$n_{\downarrow}(k)$ curves show Fermi Wave Vectors $k_F$.
      \label{fig:n_k}
  }
   \end{figure}

The energy bands $E(k)$, and the momentum distribution function of the
original fermion operators, $n(k)$, in the spiral phase, are shown in
Fig.~\ref{fig:n_k}.  An important feature to note is that only one band
crosses the Fermi energy. We have further checked that the entanglement
entropy of the system corresponds to that of spinless fermions. Because
the new electronic eigen-operators are superpositions of two different
$k$ states in the original basis, $n(k)$ shows discontinuity at four
wave-vectors. 

\begin{figure}[!h]
      \includegraphics[width=0.98\columnwidth]{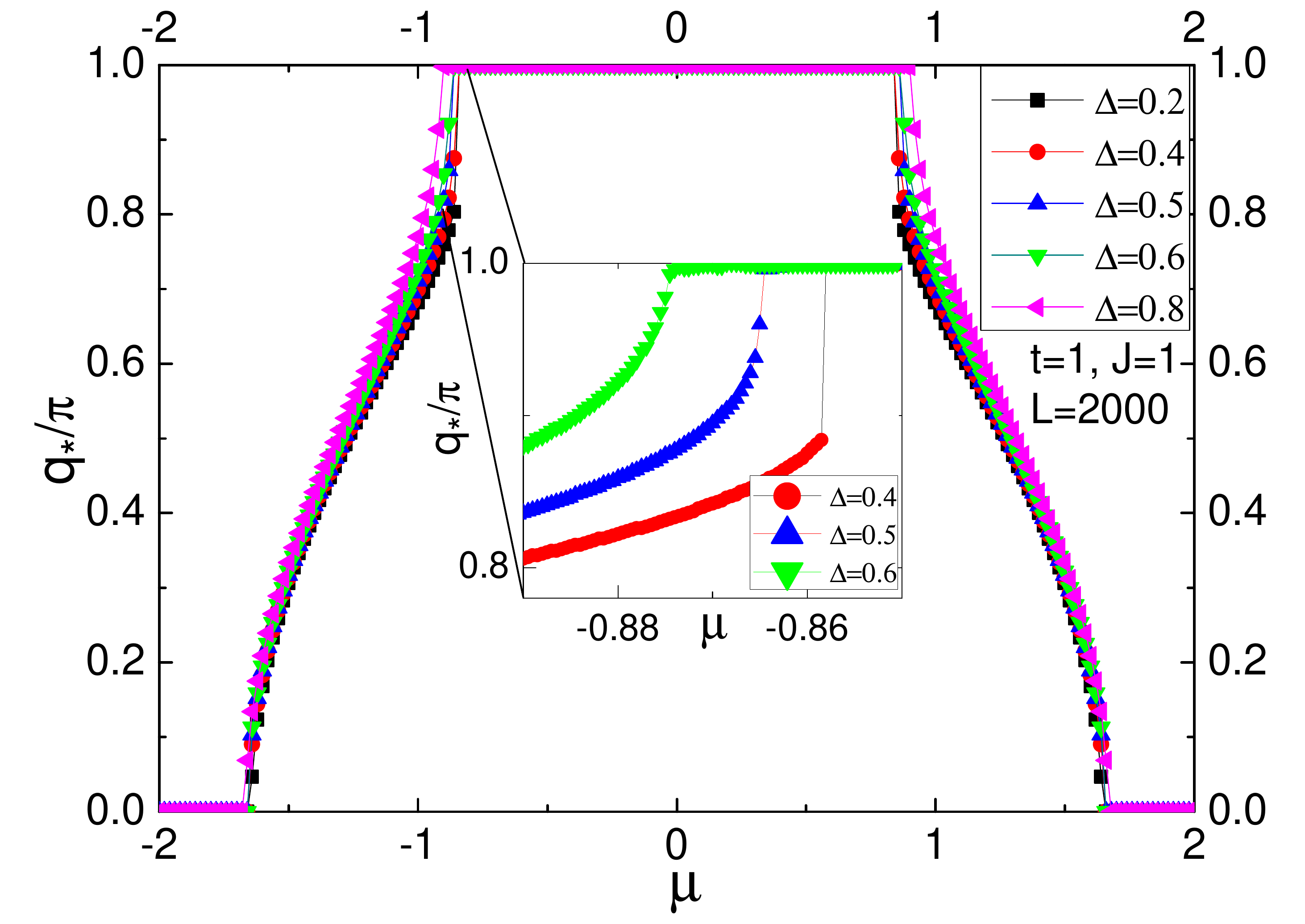} 
   \caption{
Optimal wave vector $q_*$ which minimizes the ground state energy
$E_{0}$, as a function of chemical potential $\mu$ for fixed $J=1.0$ but
different $\Delta$ values. Finite size effects are verified to be small.
\underline{Inset:} Enlarged $q_*/\pi$ vs $\mu$ graph. It has an abrupt
jump at $\Delta=0.4$ but grows up continuously at $\Delta=0.6$.
$\Delta=0.5$ is around the critical value $\Delta_c$.
      \label{fig:q_vs_mu_Delta}
  }
\end{figure}
\begin{figure}[!h]
      \includegraphics[width=0.98\columnwidth]{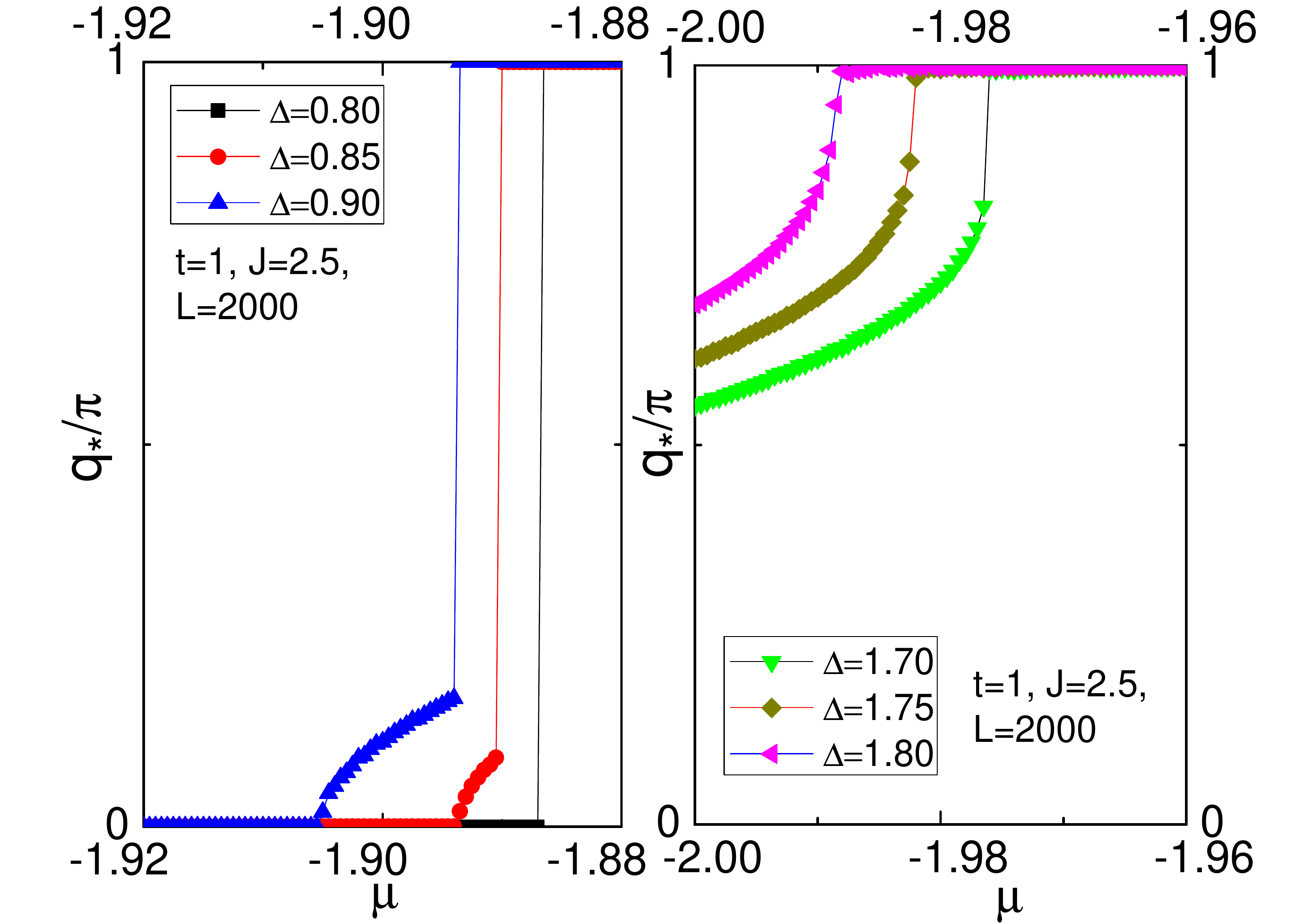} 
   \caption{
Optimal wave vector $q_*$ as a function of chemical potential $\mu$ for
fixed $J=2.5$ but different $\Delta$ values. There are two critical
values $\Delta_c$ for $J=2.5$.  \underline{Left:} Wave vector $q_*$ vs
$\mu$ for $\mu \approx \Delta^{-}_c$. This lower $\Delta^{-}_c$
distinguishes F-AF phase separation and spiral-AF phase separation.
\underline{Right:}  Wave vector $q_*$ vs $\mu$ for $\mu \approx
\Delta^{+}_c$. The upper $\Delta^{+}_c$ distinguishes spiral-AF phase
separation and no phase separation.
      \label{fig:q_vs_mu_Delta_2}
  }
\end{figure}
\begin{figure}[!h]
      \includegraphics[width=0.98\columnwidth]{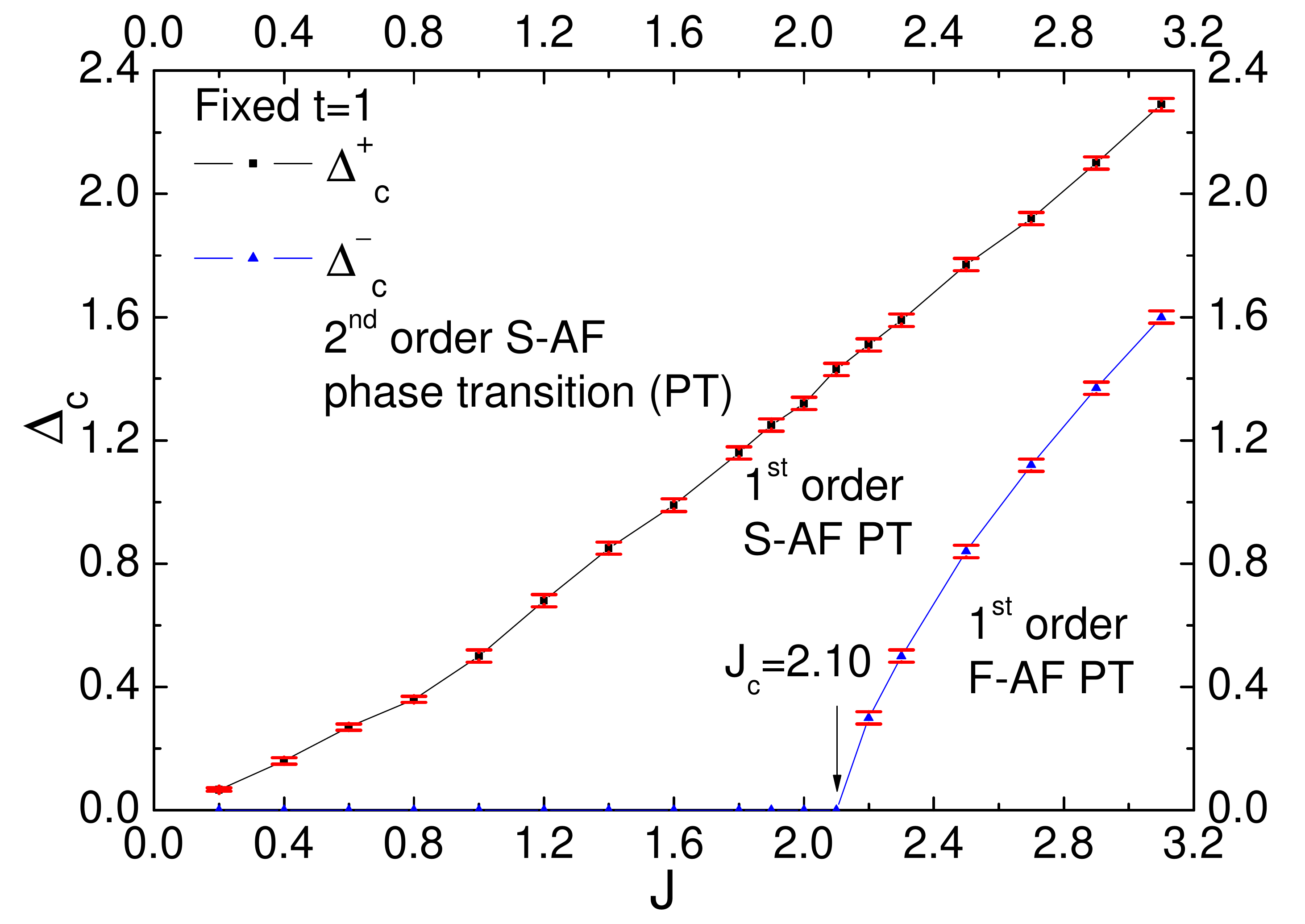} 
   \caption{
Critical value $\Delta_c$ vs exchange constant $J$. There are two
critical values, upper $\Delta^{+}_c$ and lower $\Delta^{-}_c$, for
fixed $J$. Below the critical value $J_c=2.10 \pm 0.02$, only
$\Delta^{+}_c$ is non-zero. There are three regions distinguished by
$2^{nd}$ order spiral-AF phase transition (PT), $1^{st}$ order spiral-AF
PT and $1^{st}$ order F-AF PT.
      \label{fig:Delta_critical}
  }
\end{figure}

Turning on $\Delta$ and minimizing the ground state energy $E_{0}$ (See
Eq.~\ref{eq:energy_ground}) results in the optimal ordering wave vector
$q_*$ in Fig.~\ref{fig:q_vs_mu_Delta} and
Fig.~\ref{fig:q_vs_mu_Delta_2}. $J$ mixes fermion modes of momenta $k$ and $k-q$ for a classical spin configuration of wavevector $q$, and at the same time $\Delta$ mixes $k,\sigma$ and
$-k,-\sigma$.  Together, the result is the four hybridized bands of
Eqs.~\ref{eq:energy_matrix}-\ref{eq:transformation}.  The ground state is determined by minimizing the sum of the energies of levels $\lambda_3(k)$ and $\lambda_4(k)$ which develop from $-\epsilon_{-(k-q)}$ and $-\epsilon_{-k}$, respectively. In Fig.~\ref{fig:q_vs_mu_Delta}, at $J=1.0$, the former favors a spiral phase $q_* \ne \pi$, while the latter is minimized by an AF $q_*=\pi$
We find that increasing $\Delta$ enhances the effect of $\lambda_4(k)$,
that is, makes its AF minima more pronounced than that of the
spiral minima in $\lambda_3(k)$.  This is reflected in the growth
of the size of the AF region with $\Delta$. Changing $\Delta$ from $0.2$ to $0.8$, the spiral wave vector $q_*$ increases, and results in the disappearance of spiral-AF phase separation at the critical value $\Delta_c$. We saw in
Fig.~\ref{fig:phase_diagram} that for $J \gtrsim 2t$ the spiral region
terminates and only a direct F to AF transition occurs.  An interesting
effect of the pairing term $\Delta$ is that, if it takes a sufficiently
large value, it stabilizes the spiral phase at $J \gtrsim 2t$.
Fig.~\ref{fig:q_vs_mu_Delta_2}(left) shows results for $q_*$ at $J=2.5$.
When $\Delta=0.80$, there is still a direct F ($q_*=0$) to AF
($q_*=\pi$) jump.  However at $\Delta=0.85$ a spiral phase with
intermediate $q_*$ is evident.  We define a lower $\Delta^{-}_c$ to be
the critical $\Delta$ above which the spiral is stabilized by pairing.
The jump in $q_*$ at the S to AF transition which emerges steadily
shrinks as $\Delta$ grows further.  In fact, ultimately the jump goes to
zero at an upper $\Delta^{+}_c$.  At this point the spiral to AF
transition no longer exhibits phase separation.  This behaviour is shown
in Fig.~\ref{fig:q_vs_mu_Delta_2}(right).

The shrinking of the jump in $q_*$ at the spiral to AF transition also
occurs for $J \lesssim 2t$ where the spiral phase is stable even at
$\Delta=0$ (that is, when $\Delta^{-}_c=0$).
Fig.~\ref{fig:Delta_critical} shows the critical values $\Delta^{-}_c$
and $\Delta^+_c$ as functions of $J$, and three regions with their
different types of phase transitions.

\begin{figure}[!h]
      \includegraphics[width=0.98\columnwidth]{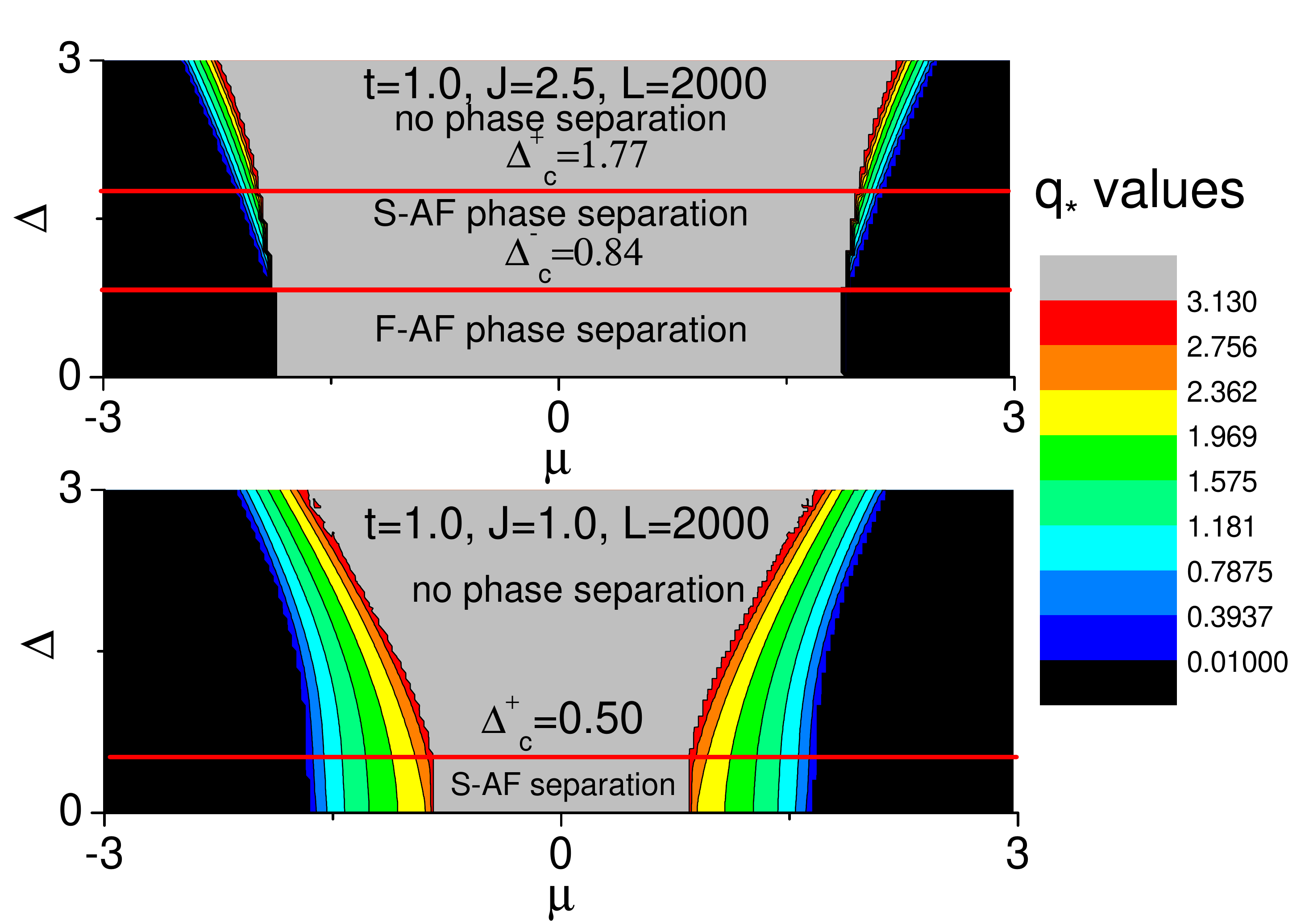} 
   \caption{\underline{Top:} Phase diagram in the chemical potential ($\mu$), pairing field ($\Delta$) plane, with fixed exchange constants ($J=2.5$). The upper red line is corresponding to upper $\Delta^{+}_c$. The lower red line is corresponding to lower $\Delta^{-}_c$. \underline{Bottom:} Phase diagram in $\mu$, $\Delta$ plane, with fixed $J=1.0$. The red line is corresponding $\Delta^{+}_c$. $\Delta^{-}_c$ is zero at $J=1.0$.
      \label{fig:phase_diagram_Delta}
  }
\end{figure}

The phase diagram of Eq.~\ref{eq:ham_Delta}, which models
a spin-fermion system in contact with an s-wave
superconductor, depends on the parameters
$\mu$, $J$ and $\Delta$.
Fig.~\ref{fig:phase_diagram} is the cut of the resulting 3D
phase diagram at $\Delta=0$.  We now explore several cuts in the
$\mu, \Delta$ plane at fixed $J$.
The resulting phase boundaries are shown for $J=1.0$
and $J=2.5$
in Fig.~\ref{fig:phase_diagram_Delta}.
(The Hamiltonian $H_{\Delta}$ is still particle-hole symmetric, so the
diagrams are symmetric about $\mu=0$.) In the bottom graph ($J=1.0$),
for which the spiral phase is stable even in the absence of pairing,
the effect of increasing $\Delta$ is to expand the stability of the AF,
and shrink the range of chemical potential for which the spiral exists.
The horizontal line shows the location of $\Delta^+_c$ where the spiral
to AF transition becomes continuous and no longer exhibits
phase separation.  The stabilization of the AF appears to onset
at $\Delta^+_c$.
In the top graph ($J=2.5$), for which the spiral is not stable
in the absence of pairing, there is a more
rich behavior.  Although the F and AF phases dominate, a
range of spiral phase arises above $\Delta^-_c$ (lower
horizontal line) and leads to spiral-AF phase separation near half-filling.
When $\Delta>\Delta^+_c$ (upper horizontal line)
phase separation disappears and the
transition from spiral phase to AF phase becomes continuous.

%%%%%%%%%%%%%%%%%%%%%%%%%%%%%%%%%%%%%%%%%%%%%%%%%%%%%%%%%%%%%%%%%%%%%%%%%%
\section{Discussion and Conclusions}
%%%%%%%%%%%%%%%%%%%%%%%%%%%%%%%%%%%%%%%%%%%%%%%%%%%%%%%%%%%%%%%%%%%%%%%%%%

In this paper, we have computed the ground state magnetic phase diagram
of one dimensional fermions on a lattice coupled to classical Heisenberg
spins.  Ferromagnetism occurs at low and high densities, and occurs only
in full polarization.  Spiral phases give way to commensurate order as
the spin-fermion coupling increases. At weak coupling $J$ the system is
thermodynamically unstable at the spiral-AF phase boundary, with
separate AF and spiral domains present  in a range of densities near
half-filling, while at strong coupling $J$, the system is
thermodynamically unstable at the F-AF phase boundary. The Maxwell
construction is verified in the phase separation. With the introduction
of $\Delta$, at weak coupling $J$, spiral-AF phase
separation survives at weak $\Delta$ but totally disappears when
$\Delta$ exceeds the critical value. At strong coupling $J$, the system
evolves from F-AF phase separation through spiral-AF phase separation to
no phase separation with the increase of $\Delta$. We should note that spiral phases may be less favorable if the overall system is higher dimensional.

A potentially exciting application of itinerant electrons interacting
with localized spins is in the context of artificially engineered
systems with magnetic atoms on the surface of a metal or a
superconductor. Indeed, the search for Majorana fermions in such hybrid
magnetic-superconducting systems is a hot
topic\cite{Alicea12,Nadjperge13,Nadjperge14} of current research.
Vazifeh {\it etal}\cite{Braunecker13,Klinovaja13,Vazifeh13} have shown
that, within a BCS treatment, when fermions are coupled to a spiral spin
configuration topological phases are robust and hence Majorana end
states should be expected\cite{Sau12}. The existence of phase separation
between spiral and antiferromagnetic (AF) states implies that, with
pairing, such Majorana fermions might move away from chain ends to the
interface between spiral and AF phases. This means that by changing the
electronic density, one may be able to move the location of the Majorana
particles. This could be helpful in the braiding of these excitations in
a network of chains. This issue deserves further consideration.

\begin{acknowledgements}
This work was supported in part by  NSF DMR-1306048
and by the Office of the President of the University of California.
\end{acknowledgements}

\end{document}